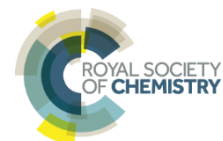

# Journal Name

## ARTICLE

# Atomically-efficient, highly stable and redox active Ce$_{0.5}$Tb$_{0.5}$O$_x$ (3% mol.)/MgO catalyst for total oxidation of methane

Juan J. Sánchez, Miguel López-Haro, Juan C. Hernández-Garrido, Ginesa Blanco, Miguel A. Cauqui, José M. Rodríguez-Izquierdo, José A. Pérez-Omil, José J. Calvino*, María P. Yeste

Redox and catalytic performance in total methane oxidation of a nanostructured ceria-terbia catalyst supported on magnesia is presented and compared to that of a pure ceria catalyst supported on MgO. The investigated material, Ce$_{0.5}$Tb$_{0.5}$O$_x$ (3% mol.)/MgO, features several remarkable properties: a quite low total molar loading of the two lanthanide elements, high reducibility as well as very high oxygen storage capacity at low temperatures and higher catalytic activity than MgO-supported CeO$_2$. In terms of lanthanide atomic content terms, the catalytic performance of Ce$_{0.5}$Tb$_{0.5}$O$_x$ (3% mol.)/MgO largely improves that of bulk type ceria and ceria-magnesia solid solutions. Such behavior implies a proper optimization in the usage of the lanthanide elements. A second contribution to atomic economy in the catalyst design relates to the fact that the novel formulation demonstrates a good stability in the redox and catalytic performance against very high temperature treatments.  An investigation of the structure of both the fresh and high temperature-aged catalyst at atomic scale, by means of complementary aberration corrected microscopy techniques, reveals the occurrence of a variety of highly dispersed, exotic, lanthanide-containing nanostructures, which span from isolated, atomically dispersed, Ln species, to nanometer-sized CeTbO$_{2-x}$ patches, extended CeTbO$_{2-x}$ bilayers and 3D CeTbO$_{2-x}$ nanoparticles. Nanoanalytical results evidence mixing of the two lanthanides at atomic levels in these nanostructures. The combined effect of nanostructuring, mixing of the lanthanides at atomic level and interaction with the MgO oxide are at the roots of the improvement in functional, redox and catalytic properties of the novel Ce$_{0.5}$Tb$_{0.5}$O$_x$ (3% mol.)/MgO catalyst.

## Introduction

Prompted by the large increase in the availability of natural gas and its lower impact on CO$_2$ and NOx emissions, the production of energy by the combustion of methane has become a field of growing interest.[1–3] Thus, natural gas is considered an alternative to gasoline as a fuel for internal combustion engines,[4] in the so-called Natural Gas Vehicles (NGVs), as well as for direct electrochemical conversion in SOFCs. From a pollutant abatement point of view, combustion of methane is a model reaction to check material´s performance in catalytic applications related to the removal of hydrocarbons from automobile exhaust gases, e.g. in TWCs,[5] while it is a central process in the removal of residual levels of this green-house gas in the emissions from NGVs. The expected increase in the rate of incorporation of this type of vehicles (around 20% annually)[6] combined with the much larger global warming potential of CH$_4$, in comparison to CO$_2$,[7] contribute also to the renewed interest in the process of catalytic total combustion of methane.

Supported noble metals, mostly Pt and Pd, have been generally used as catalysts for this reaction,[8,9] but Palladium is reported to show the highest low temperature activity.[10–16] The wealth of research on supported Pd catalysts points out to the key role of the oxidized forms of the metal[11] as well as to the impact of redox promoters, such as CeO$_2$, on catalytic performance of Pd.[17,18] Thus, due to the intrinsic high Oxygen Storage Capacity (OSC) of CeO$_2$,[19] oxygen transfer in Pd/CeO$_2$ is thought to sustain an oxidized state of the metal nanoparticles under temperature conditions at which decomposition of the labile Pd-O bond is favored. Moreover, a key role of the metal interface sites in this process has been clearly identified.[20]

From these observations, high oxygen exchange capabilities in a wide range of temperatures,[21] including low temperatures; high stability of the redox response at high temperatures and capability to provide an extended as possible contact with a supported metal phase appear as key features in the design of catalytic supports for this reaction.[22] Likewise, from an atomic economy point of view, the optimization in the use of the critical raw materials employed in their formulation, as it is clearly the case of Rare Earths,[23] should also guide the development of these novel supports.

The dispersion of CeO$_2$ on the surface of a carrier oxide and further nanostructuration of the resulting supported ceria into the form of highly dispersed phases has recently proven as an efficient route to obtain materials depicting very high Oxygen Storage Capacities (OSCs), even at low temperatures, as well as outstanding stability in the redox response against high temperature treatments. This has been achieved in catalysts supported on $ZrO_2$, YSZ and MgO featuring total ceria contents in the order of just 10% mol. or, even, lower.[24–28] In the particular case of a 6%(mol.) $CeO_2$/MgO catalyst, the supported ceria phase was present in the form of a mixture of very highly dispersed nanostructures, spanning from ceria nanosized patches, extended atomic bilayers, to isolated Ce atoms. Such highly dispersed ceria structures are present even after treatments at temperatures as high as 900ºC, coexisting with 3D $CeO_2$ nanometer-sized particles. They represent an optimal situation to interact with a secondary metallic phase added in an ulterior synthesis step. Additionally, the enhanced OSC behavior observed in the systems prepared on MgO can fulfil the requirement of acting as an efficient active oxygen source. These works clearly evidence that nanostructuration in the form of atomically thin nanostructures (nano-patches, bilayers or isolated atoms) provide a route to imprint enhanced redox properties in highly dispersed ceria species. Recent contributions in which powder type catalysts are prepared by Atomic Layer Deposition (ALD) have in fact evidenced the benefits in catalytic performance of phases structured as supported thin films.[29–31] Nevertheless, the synthetic approach used in [24-28] involve the use of conventional wet impregnation techniques, which represents a large advantage from the synthetic point of view.

At this point, two other questions are worth mentioning. The first refers to the beneficial effect of basicity on limiting both carbon deposition and sulfur poisoning,[32,33] two major deactivation mechanisms operating in methane combustion.

The second one is related to the positive effects on $CeO_2$ functional properties induced by doping with transition metals or other lanthanides. Thus, improvements in the redox response after incorporation of these elements into the host fluorite structure of $CeO_2$ have been clearly identified and investigated in depth in the literature.[34–38] Likewise, promotional effects on methane oxidation activity after modification with La and Zr or Ca and Nd have been attributed to changes in both reducibility and sintering resistance at high temperatures.[39,40] More recently, in a comparative study among $CeO_2$, $Ce_{0.8}Zr_{0.2}O_2$ and $Ce_{0.8}Hf_{0.2}O_2$ Zamar et al.[41] found that the incorporation of the isovalent $Zr^{4+}$ and $Hf^{4+}$ ions results in an enhanced methane oxidation activity, which was enabled by the improved OSC of the resulting defect-fluorite structure.

On the basis of all this previous knowledge, in this contribution we explore the preparation by conventional, widely available, wet chemistry routes, of a MgO supported ceria-terbia mixed oxide catalyst depicting a very low lanthanide loading. The a-priori expectation is that in this reformulated catalyst, $Ce_{0.5}Tb_{0.5}O_x$ (3% mol.)/MgO, a highly dispersed lanthanide phase is kept and that the interaction at atomic level between Ce and Tb results in an improvement in redox and methane oxidation activity, with respect to both pure $CeO_2$ and $CeO_2$/MgO.

The presence of the mixed ceria-terbia component as a highly dispersed phase with improved redox properties is investigated in depth by the combined use of a variety of advanced electron microscopy and macroscopic (XRD, XPS, TPR) techniques since, as commented above, these are features of large interest in the behavior of the material as catalytic support for this reaction. Given that the methane combustion is a very exothermic reaction, special attention has been also paid to the structural and functional stability of the catalyst against high temperature treatments.

## Experimental

The MgO used as support was obtained by calcination in air at 450ºC for three hours of a $Mg(OH)_2$ commercial sample from Alfa Aesar. The $Ce_{0.5}Tb_{0.5}O_x$ (3% mol.)/MgO catalyst was prepared by incipient wetness impregnation using an acetone solution containing equimolar concentrations of $Ce(NO_3)_3 \cdot 6H_2O$ and $Tb(NO_3)_3 \cdot 6H_2O$. Acetone was used as solvent, instead of water, in order to avoid the uncontrolled dissolution of MgO by acid-base reaction with the impregnating solution. Let us recall at this respect that MgO is a basic oxide and that aqueous solutions of metals are highly acidic. To avoid the fast evaporation of the solvent during the deposition of the lanthanides, the impregnation step was carried out at -10ºC in a single step. The concentration of the solution containing the two lanthanides was adjusted to obtain a final molar loading of 3%.

After impregnation, the catalyst was dried in oven at 105ºC overnight, grounded in an agate mortar, sieved through a 75 ! m mesh and further calcined under air in an oven at 500ºC. A 5ºC/min heating ramp was used for the calcination step. Temperature was kept for 1h at 500ºC and then the solid was cooled down under air to room temperature and finally stored in a desiccator. This sample will be referred in the following as $Ce_{0.5}Tb_{0.5}O_x$ (3% mol.)/MgO.

To study the influence of high temperatures treatments on textural, compositional, structural and functional properties, a portion of the $Ce_{0.5}Tb_{0.5}O_x$ (3% mol.)/MgO catalyst was submitted to a redox aging cycle consisting of the following steps:

Severe Reduction or SR: heating under a flow of $H_2$(5%)/Ar (60 $cm^3 \cdot min^{-1}$) from room temperature up to 950ºC, at a heating rate of 10 ºC $min^{-1}$ and followed by 2 h of isothermal treatment at 950ºC. Next, the gas flow was switched to He (60 $cm^3$ $min^{-1}$), for 1 h, and the sample was cooled down to 25ºC under inert gas flow (60 $cm^3 \cdot min^{-1}$).

After this SR step, the aging cycle was closed by applying a mild re-oxidation, MO, routine. To prevent overheating of the reduced mixed oxides, this part of the cycle started with a very low temperature re-oxidization at -80ºC, by flowing an $O_2$(5%)/He mixture (60 $cm^3 \cdot min^{-1}$) for 1h over the catalyst. The sample was then allowed to freely warm up to room temperature under the same flow of $O_2$(5%)/He (60 $cm^3 \cdot min^{-1}$) and then from room temperature up to 500ºC using a heating ramp of 10ºC/min. After reaching this temperature, the oxidation treatment was prolonged for 1 h. Finally, the sample was cooled down under the same atmosphere to room temperature. In the following this catalyst will be referred as $Ce_{0.5}Tb_{0.5}O_x$ (3% mol.)/MgO SRMO.

Textural properties of the samples were determined by $N_2$ adsorption at -196 ºC using a Quantachrome Autosorb iQ automatic device. Before measurement, samples were pre-evacuated at 200ºC for 2 h.

Inductively coupled plasma-atomic emission spectrometry (ICP-AES) was employed to determine the Ce and Tb loadings of the catalyst. X-ray diffraction (XRD) analyses of the catalysts were carried out using a Bruker diffractometer Model D8 ADVANCE operated at 40 kV and 40 mA employing Cu Kα radiation.

Temperature Programmed Reduction experiments under hydrogen, $H_2$-TPR-MS, were performed in an experimental device coupled to a Pfeiffer, model Thermostar QME-200-D-35614, quadrupole mass spectrometer using 200 mg of sample, a 5% $H_2$/Ar flow rate of 60 cm$^3$ min$^{-1}$ and a heating ramp of 10 ºC min$^{-1}$. Prior to all the $H_2$-TPR runs, the samples were cleaned by heating up to 500ºC under 5% $O_2$/He flowing at 60 cm$^3$ min$^{-1}$, at a heating rate of 10 ºC min$^{-1}$; then, they were kept for 1 h at this temperature and further cooled down to 150ºC under the flow of diluted oxygen and finally to 25ºC in He. The results are presented in the form of water evolution (mass/charge ratio=18) *vs* temperature, during the reduction process.

For a quantitative analysis of the reduction process, $H_2$-TPR was also performed using the Thermal Conductivity Detector (TCD) of an Autochem Micromeritics apparatus. The amount of sample routinely used in these experiments was 30 mg, the 5% $H_2$/Ar flow rate was 25 cm$^3$ min$^{-1}$, and the heating ramp was 10°C min$^{-1}$. Prior to all the TPR runs the samples were cleaned by heating them under flowing He at 25 cm$^3$ min$^{-1}$, at a heating rate of 10°C min$^{-1}$, up to 500°C. Then, they were kept for 1h at this temperature and cooled down to 150ºC under the flow of diluted oxygen and finally to 25ºC in He.

Ultimate Oxygen Storage Capacity (OSC) values were obtained from oxygen volumetric chemisorption experiments. The isotherms were recorded at 200ºC on a Micromeritics ASAP 2020 instrument in the oxygen partial pressure interval 0–300 Torr. The samples (400 mg) were pre-reduced by heating in a flow of 5% $H_2$/Ar (60 cm$^3$ min$^{-1}$) at 10 ºC min$^{-1}$, from 25ºC up to the selected reduction temperature ($T_{redn}$); they were kept at $T_{redn}$ for 1 h under flowing 5% $H_2$/Ar; then they were evacuated for 1 h (residual pressure < 1.10$^{-6}$ Torr) at $T_{redn}$ or 500ºC if $T_{redn}$ was lower than 500ºC and then finally cooled down to 200ºC under high vacuum. These evacuation conditions ensure the elimination of any significant amount of hydrogen chemisorbed on the oxides.

Samples for TEM/STEM analysis were obtained by directly depositing a tiny portion of the nanocrystal powders onto holey carbon-coated TEM grids at room temperature in order to avoid contact with any solvent which could cause contamination during STEM observation. TEM/STEM characterization was performed using a FEI Titan Cubed Themis 60-300 microscope, equipped with a monochromator, operated at 200 kV and 80 kV. For HR-STEM-HAADF imaging, a semi convergence angle of 20 mrad was used together with a camera length of 115 mm. This microscope was equipped with a high efficiency XEDS ChemiSTEM system implementing 4 windowless SDD detectors.[42] Quantification of the XEDS data was performed using the Bruker ESPRIT software. Very high spatial resolution EELS experiments were performed working in the spectrum imaging (SI) mode[43], which allows the correlation of analytical and structural information of selected regions of the material under study. In this technique, the EELS and HAADF signals are collected simultaneously while the electron beam is scanned across the selected area of the sample. The SI experiments were acquired in Dual EELS mode using an energy dispersion of 0.25 eV, 80 pA probe current and 100 ms acquisition time per EELS spectrum. In the Dual EELS mode, the zero-loss region is recorded simultaneously with the core-loss signal of the element(s) of interest, which allows a very precise determination of the absolute value of the energies at which the core-loss edges are appearing in the experiment. The latter is a key aspect when the determination of oxidation states of the elements is pursued in the EELS experiment. In our case, Ce and Tb $M_{4,5}$ (885-905 eV, 1241-1275 eV, respectively) elemental and oxidation state maps were built after removing the background from raw data, using a power law model and a window width of 25 eV. HREM images were recorded at 80 kV by exciting the momochromator up to 0.4, which allowed obtaining an energy resolution of 200meV. Afterwards, the aberration of the objective lens was corrected up to fourth-order using the Zemlin tableau.

X-ray Photoelectron Spectra were obtained using a Kratos Axis Ultra DLD instrument, and recorded with monochromatic Al Kα radiation (1486.6 eV). The instrument was operated in the fixed analysis transmission mode (FAT), using pass energy of 20 eV. The Kratos coaxial charge neutralization system was used to compensate charging effects, and the binding energy scale was calibrated with respect to binding energy (BE) scale was calibrated with respect to the highest BE peak for Ce 3d core level (u''') at 917.0 eV.[44] CasaXPS software (version 2.3.19rev1.1m), Casa Software Ltd., Devon, UK, 2017) was used for spectra processing.

## Results and discussion

### Macroscopic Characterization

The exact molar content in each lanthanide of the starting magnesia supported ceria-terbia catalyst was determined by Inductively Coupled Plasma-Atomic Emission Spectroscopy (ICP-AES). Values of 1.6% Ce and 1.5% Tb were found, which very closely correspond to the nominal values. With these loadings, Ce represents 53% of the total lanthanide content but, for simplicity, the prepared catalyst will be referred as to $Ce_{0.5}Tb_{0.5}O_x$(3% mol.)/MgO. An *x* subscript is placed on oxygen, since the actual reduction state of the lanthanides in the as-prepared sample is not known before detailed characterization. Though no compositional change was expected during the SRMO redox aging cycle, the molar loadings were confirmed by ICP-AES after such treatment.

Table 1 gathers the surface area of the two prepared catalysts. Note that the initial sample depicts a high surface area, which largely drops after the SRMO cycle. This textural change is also observed in the corresponding $N_2$-isotherms, Figure S1.

Table 1. Textural characteristics of the prepared catalysts.

| SAMPLE | $S_{BET}$ (m$^2$ g$^{-1}$) |
|---|---|
| Ce$_{0.5}$Tb$_{0.5}$O$_x$(3% mol.)/MgO | 90 |
| Ce$_{0.5}$Tb$_{0.5}$O$_x$(3% mol.)/MgO SRMO | 27 |

The structure was first investigated at macroscopic level by means of X-Ray Diffraction (XRD). The XRD diagram of the Ce$_{0.5}$Tb$_{0.5}$O$_x$(3% mol.)/MgO catalyst, Figure S2(a), contains wide diffraction peaks corresponding to a component with fluorite-type structure, in addition to the peaks characteristic of periclase MgO. Though the reflections of the fluorite phase are rather broad, a comparison of their position with those of CeO$_2$, TbO$_2$ and Tb$_2$O$_3$ indicates that they lie between those of CeO$_2$ and Tb$_2$O$_3$, but far from those of TbO$_2$. This points out that Tb is present in this catalyst mostly as Tb$^{3+}$ ions. Let´s recall at this respect that the size of Ce$^{4+}$ (97 pm) is only slightly smaller than that of Tb$^{3+}$ (98 pm), but larger than that of Tb$^{4+}$ (88 pm), due to the lanthanide contraction effect. The crystallite size of the fluorite and MgO phases, determined by the Scherrer method, were 7 nm and 14 nm, respectively.

After the SRMO treatment no new phase appears in the XRD diagram, Figure S2(b). The peaks of the fluorite phase are now better observed to be closer to those of Tb$_2$O$_3$. Narrowing of all the peaks indicate a bigger crystallite for the supported phase, 9 nm, and particularly for the MgO support, 22 nm.

The surface composition of the catalysts was investigated by means of XPS, Figure 1 and Table 2. After quantitative analysis of the experimental spectra and concentrating first on the as-prepared catalyst, we should highlight the following features:

(1) The Ce/Tb molar ratio, 1.1, as determined from the integrated intensities of the Ce 4d and Tb 4d signals, is exactly that expected from the ICP determined values (1.6 / 1.5 = 1.1).

(2) The value of the (Ce 4d + Tb 4d)/Mg 2s intensity ratio, 0.27, is much higher than corresponding to the bulk value, 0.031 (=3/97), which clearly evidences that the two lanthanides concentrate on the surface. To understand this question, it is important to take into account that the values of the inelastic mean free path (IMFP) of Ce 4d, Tb 4d and Mg 2s photoelectrons are 2.18 nm, 2.13 nm and 2.21 nm and also that about 95% of the XPS signal comes from roughly 3 times the IMPF, i.e. ≈ 6 nm. The same is observed when the (Ce 3d + Tb 3d)/Mg 1s ratio, 0.67, is compared with the average composition of the catalyst. Since the depth of analysis in the latter case is smaller, roughly less than 3 nm, the much larger deviation between the XPS and average ratios clearly points out to a concentration of the lanthanides in the very first atomic layers of the surface.

(3) The analysis of the oxidation state of the lanthanides indicates that no Tb$^{4+}$ could be detected and that the fraction of Ce$^{3+}$ amounts to 15%, Figure 1.

(4) The analysis of the C 1s signal, Figure S3, indicates the presence of residual carbonates. The intensity ratio between the C 1s signal attributable to carbonates and that of Mg 1s amounts to 0.37.

After SRMO, XPS evidences a slight increase in the Ce 4d/Tb 4d ratio up to 1.3±0.1, showing that there are no major changes in the lanthanide distribution in the mixed oxide crystals, at surface level. A decrease of the (Ce 4d + Tb 4d)/Mg 2s and (Ce 3d + Tb 4d)/Mg 1s ratios to 0.13 and 0.25, respectively, was also observed. This is consistent with the growth of the crystallites of the fluorite phase detected by XRD. Since the surface area of the catalyst decreases largely after the redox cycle, part of the fluorite phase could have also become buried at the interfaces between merging MgO crystallites. This could represent another contribution to the observed changes in the total lanthanide to support intensity ratios.

The percentage of Ce$^{3+}$ species increases after SRMO slightly up to 23% and Tb$^{3+}$ contribution remains unchanged (~100%). As expected, the fraction of carbonates decreases to a C 1s/Mg 1s ratio of 0.14.

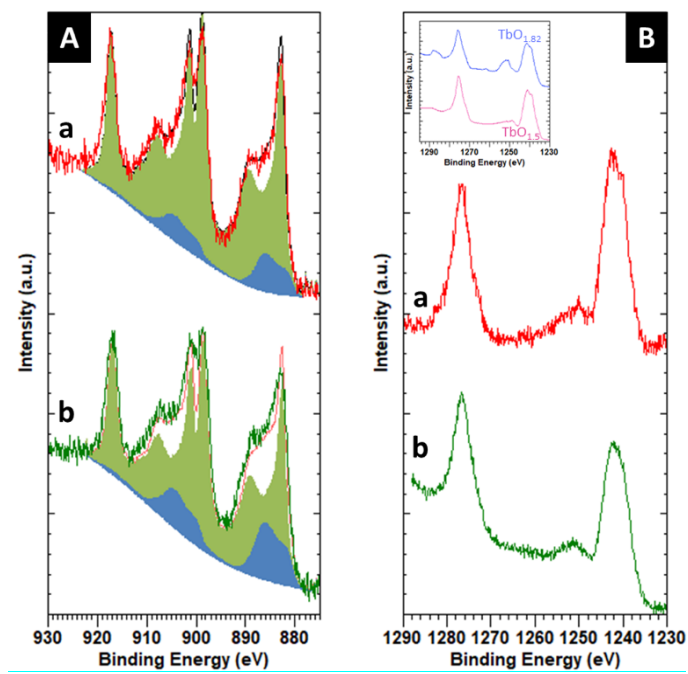

Figure 1. Ce3d (A) and Tb3d (B) XPS signals of Ce$_{0.5}$Tb$_{0.5}$O$_x$(3% mol.)/MgO (a) and Ce$_{0.5}$Tb$_{0.5}$O$_x$(3% mol.)/MgO SRMO (b). Ce 3d peak decomposition using reference Ce$^{4+}$ (green) and Ce$^{3+}$ (blue) spectra. Inset in (B) shows reference spectra for TbO$_{1.82}$ (mixed Tb$^{4+}$ and Tb$^{3+}$) and TbO$_{1.5}$ (pure Tb$^{3+}$).

### Redox Properties

The reducibility of the catalysts was investigated by means of Temperature Programmed Reduction under hydrogen experiments followed by mass spectrometry (H$_2$-TPR-MS), Figure 2.

Table 2. XPS quantification of the prepared catalysts.

| SAMPLE | Ce 4d/Tb 4d | (Ce 4d+Tb 4d) / Mg 2s | (Ce 3d+Tb 3d) / Mg 1s | C 1s (carbonate) / Mg 1s |
|---|---|---|---|---|
| Ce$_{0.5}$Tb$_{0.5}$O$_x$(3% mol.)/MgO | 1.1±0.1 | 0.27 | 0.67 | 0.37 |
| Ce$_{0.5}$Tb$_{0.5}$O$_x$(3% mol.)/MgO SRMO | 1.3±0.1 | 0.13 | 0.25 | 0.14 |

# Journal Name

## ARTICLE

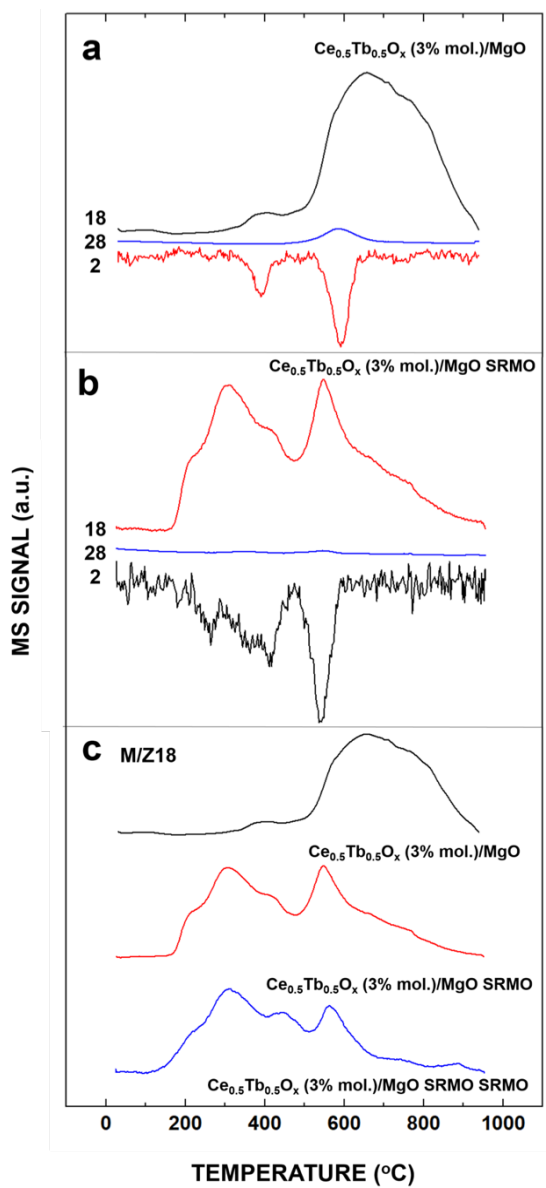

Figure 2. (a) Evolution of (m/z=18, 28 and 2) during the H$_2$-TPR-MS analyses of Ce$_{0.5}$Tb$_{0.5}$O$_x$ (3% mol.)/MgO (b) Evolution of (m/z=18, 28 and 2) during the H$_2$-TPR analyses of Ce$_{0.5}$Tb$_{0.5}$O$_x$ (3% mol.)/MgO SRMO (c) Evolution of water (m/z=18) during consecutive H$_2$-TPR analyses of the Ce$_{0.5}$Tb$_{0.5}$O$_x$(3% mol.)/MgO catalyst.

In the case of the Ce$_{0.5}$Tb$_{0.5}$O$_x$(3% mol.)/MgO catalyst, the traces of the mass/charge (m/z) ratios corresponding to the evolution of H$_2$O and CO as well as the consumption of H$_2$, plotted in Figure 2(a), indicate the occurrence of two major reduction events, one peaking at 400ºC and a second one with maximum at roughly 600ºC. The first peak involves both H$_2$ consumption and H$_2$O evolution, whereas in the second evolution of both H$_2$O and

CO takes place. Therefore, the first event can be assigned to oxygen abstraction from the catalyst but the second one contains, additionally, a contribution of the reduction of residual carbonates. The trace of water shows a third, broad, signal, covering the 700ºC-900ºC range, which overlaps with the signal at 600ºC. A Temperature Programmed Oxidation (TPO) experiment carried out on the Mg(OH)$_2$ used as precursor of the MgO support, Figure S4, evidences water evolution in this temperature range. If we recall that MgO was obtained by calcination at 450ºC of the hydroxide, it seems clear that the high temperature H$_2$O signal in the H$_2$-TPR-MS does not correspond to a reduction event but, instead, to the evolution of residual OH groups in the MgO support.

Noticeably, after SRMO, the H$_2$-TPR-MS shows, Figure 2(b), in addition to the two reduction events observed in the fresh catalyst, an additional H$_2$ consumption in the very low temperature range, between 200ºC and 400ºC. This result does not only indicate a large stability of the redox performance of the prepared catalyst but, also, an improvement of its reducibility after being submitted to a high temperature redox cycle. The stability is further confirmed by the fact that a second H$_2$-TPR-MS performed on the Ce$_{0.5}$Tb$_{0.5}$O$_x$ (3% mol.)/MgO-SRMO, Figure 2(c), results in a reducibility response which is nearly the same as that observed on the starting Ce$_{0.5}$Tb$_{0.5}$O$_x$ (3% mol.)/MgO-SRMO catalyst. Therefore, cycling the material at high temperature does not deteriorate its redox response but, contrarily, improves its reducibility at low temperatures.

To estimate the total percentage of reduction of the lanthanides, H$_2$-TPR-TCD experiments were recorded. Quantification of the corresponding diagrams, including the whole temperature range, resulted in the H2 consumption values included in Table 3. The first entry of this Table, 345 µmoles/g, corresponds to the amount of hydrogen that would be required for the reduction of the two lanthanides, in case they were both present in the catalyst only as Ln$^{4+}$ species. The 244 µmoles/g consumed by Ce$_{0.5}$Tb$_{0.5}$O$_x$ (3% mol.)/MgO therefore indicates that a fraction of the lanthanides was already reduced, in good agreement with the XPS observations. In fact, this quantity can be fully explained considering the Ce3+ and Tb$^{3+}$ percentages determined by XPS (15% and 100%, respectively) and a contribution due to the reduction of 5% wt. of carbonates, as revealed by the H$_2$-TPR-MS experiments. This later value is also in good agreement with the results of the analysis by ICP. The consumption of hydrogen in the second H$_2$-TPR-TCD cycle, which is in fact the first H$_2$-TPR-TCD of the Ce$_{0.5}$Tb$_{0.5}$O$_x$ (3% mol.)/MgO-SRMO, amounts to 157 µmoles/g. This corresponds to the reduction of Ce$^{4+}$ in an amount corresponding to 85% of the total Ce content, which is close to the fraction (77% Ce$^{4+}$) determined by XPS.

Table 3. Hydrogen consumption in H₂-TPR-TCD of the prepared catalysts

| SAMPLE | μmol H₂/g-cat |
|---|---|
| Theoretical H₂ consumption | 345 |
| First H₂-TPR-TCD | 249 |
| Second H₂-TPR-TCD | 157 |
| Third H₂-TPR-TCD | 154 |

Table 4. Evolution of OSC with reduction temperature. *

| SAMPLE | 400°C | 500°C | 700°C |
|---|---|---|---|
| CeO₂ (6% mol.)/MgO | 18 | 44 | 67 |
| CeO₂ (6% mol.)/MgO SRMO | 12 | 54 | 70 |
| Ce₀.₅Tb₀.₅Oₓ (3% mol.)/MgO | 49 | 52 | 41 |
| Ce₀.₅Tb₀.₅Oₓ (3% mol.)/MgO SRMO | 27 | 39 | 47 |

* expressed as %Ln³⁺

Finally, after the third H₂-TPR-TCD run on the fresh catalysts, i.e. second run on the SRMO one, the consumption of hydrogen does not change significantly, which clearly illustrates the high stability of the redox response of the prepared catalyst against redox cycling.

The Oxygen Storage Capacity (OSC) of the two catalysts at increasing temperatures was also measured, Table 4. The results obtained on a CeO₂ (6% mol.)/MgO catalyst are included for comparison. In order to highlight the efficiency in the usage of the lanthanide elements, the oxygen consumption values have been translated into percentage of reduction referred to the total content in lanthanides. Thus, the 49% value determined for the Ce₀.₅Tb₀.₅Oₓ (3% mol.)/MgO sample at 400°C means that the treatment with the H₂(5%)/Ar mixture at this temperature has been able to reduce such percentage of the total lanthanide content. Since Tb is fully reduced in the starting catalyst, according to XPS and H₂-TPR data, and represents 47% of the total lanthanide content, this OSC value means that 92% of Ce present in the catalyst gets reduced and that the total Ln³⁺ content in the catalyst reduced at 400°C amounts to 96%.

Note how the incorporation of Tb leads to a large increase of the reducibility of the catalyst, particularly at low temperatures. In fact, only 18% of Ce transforms into Ce³⁺ in the CeO₂ (6% mol.)/MgO catalyst after reduction at 400°C whereas, as just commented, in the Tb-modified catalyst, this quantity increases up to 92%. At 500°C, 96% reduction of Ce takes place in this catalyst, whereas in the pure ceria supported one reduction only involves 44% of this lanthanide. The presence, as evidenced by XPS, of a fraction of Ce³⁺ in the as-prepared catalyst explains a value slightly below 100% Ce content. Even at 700°C, reduction only affects 67% of Ce in CeO₂ (6% mol.)/MgO. At this temperature, 700°C, the OSC of the Ce₀.₅Tb₀.₅Oₓ (3% mol.)/MgO catalyst decreases only slightly, maybe due to sintering of both the supported phase and the support.

The SRMO treatment deteriorates the OSC of the Ce₀.₅Tb₀.₅Oₓ (3% mol.)/MgO catalyst at 400°C and 500°C, but still the values observed for the reduction of Ce remain above those of the catalyst based on pure ceria submitted to the same treatment, CeO₂ (6% mol.)/MgO-SRMO. Thus, at 400°C 12% of Ce is reduced in this later catalyst, while the value increases up to 51% in Ce₀.₅Tb₀.₅Oₓ (3% mol.)/MgO-SRMO. Likewise, 74% and 89% reduction of Ce takes place in Ce₀.₅Tb₀.₅Oₓ (3% mol.)/MgO-SRMO at 500°C and 700°C; both values standing over those observed in the pure ceria based one. At 700°C, OSC improves slightly and, in the case of the Tb-modified catalyst, reaches a value which represents close to full reduction of all Ce⁴⁺ present in the catalyst.

Summarizing, the redox studies carried out clearly evidence an improvement in the reducibility of the catalyst after the incorporation of Tb in the formulation. This improvement translates not only into a higher OSC at low temperatures but also in a much better reducibility of Ce in the whole temperature range with respect to the pure CeO₂/MgO catalyst, both before and after SRMO.

**Atomic Scale Structural and Chemical Characterization**

An in-depth analysis of the structure of the prepared catalyst both before and after the SRMO treatment was performed by means of a set of complementary Transmission and Scanning-Transmission Electron Microscopy techniques using a state-of-the-art, double aberration corrected and monochromated, TEM/STEM microscope. The goal of such study was, particularly, determining the type of nanostructures into which the supported phase was present, its dispersion state and interactions with the underlying MgO support.

Concentrating first on the as-prepared, Ce₀.₅Tb₀.₅Oₓ (3% mol.)/MgO catalyst, Figure 3 shows two representative High Resolution High Angle Annular Dark Field STEM (HR-HAADF-STEM) images which illustrate the structure of this sample. Since the contrasts in these images are directly related to, roughly, the square of the atomic number of the atoms under the beam, the supported phase is easily recognized (Z_Ce=58, Z_Tb=65) and differentiated from the MgO support (Z_Mg=12), as the higher intensity areas. Note that the ceria-terbia component is in the form of nanometer-sized, 2-3 nm, crystallites. The analysis of the contrasts in the frequency space clearly indicates that these crystallites correspond to a fluorite structure. In the diffractogram shown as inset in Figure 3(a), the {002} and {220} spacing values characteristic of this phase are identified. The different crystallites are not oriented between each other, as reveals the appearance of reflection rings of both MgO and the fluorite phase when in the diffractograms of large areas, Figure S5. Another important feature is the presence of a large number of isolated atoms. These are clearly observed as high intensity spots all over the imaged areas. At some locations, these highly dispersed species assemble into cluster type structures. Regarding the MgO support, note that this is constituted by an agglomeration of pretty small, crystallites. These crystallites appear also randomly oriented to each other, Figure S5. The analysis of the lattice of these crystallites in the Fourier space shows the features characteristic of MgO (see inset in Figure 3(a) and Figure S5).

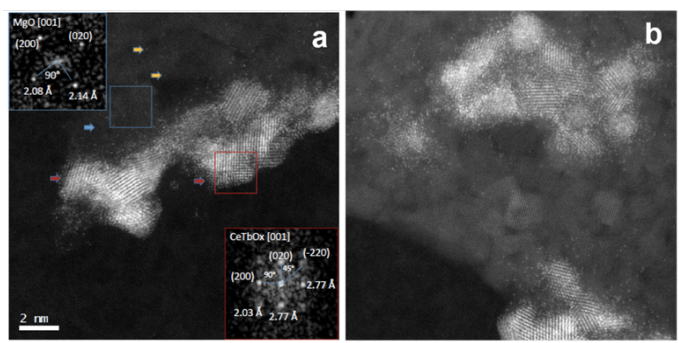

Figure 3. Representative HR-HAADF-STEM images of the Ce$_{0.5}$Tb$_{0.5}$O$_x$ (3% mol.)/MgO catalyst.

To determine the spatial distribution of the different elements in the catalyst, STEM-EDS chemical maps were acquired. Figure 4 shows representative results of this nanoanalytical study. In the HAADF-STEM image of the analyzed area, Figure 4(a), a collection of crystallites of the fluorite, lanthanide containing phase is clearly observed on the edge of the MgO support, as confirmed from the Ce, Tb and Mg maps, Figures 4(b)-(d). According to these maps, the two lanthanides are mixed at atomic level in the nanosized particles. The quantification of the spectrum indicates Ce and Tb molar percentages of 42% and 58%, values close to the average composition.

Since stability is a quite relevant issue, the structure of the catalyst after SRMO was also investigated in detail. Figure 5, shows HAADF-STEM and HREM images of the Ce$_{0.5}$Tb$_{0.5}$O$_x$ (3% mol.)/MgO-SRMO sample. In both imaging modes the presence of a large number of contrasts in the form of individual atoms is observed. In the HAADF-STEM images two different high intensity levels are observed (inset in Figure 5(b)) as it is also the case, in terms of two different low intensity levels, in the HREM image (inset in Figure 5(d)). This suggests that a fraction of both Ce and Tb is present in the catalyst in the form of isolated atomic species. Moreover, the HAADF-STEM image clearly indicates that these isolated species are mostly located on top of the contrasts of the MgO lattice planes. It is also evident that, against expectation, the fraction of these isolated species has increased with respect to the as-prepared catalyst, in such a way that at some locations they merge into cluster or patch-like structures, Figure 5(a).

In the HAADF-STEM image recorded in edge-on view shown in Figure 6(a) another type of nanostructure is observed, extended bilayers. Measurements of the lattice spacing in images, like that in Figure 6(a), on intensity profiles recorded along the direction perpendicular to the interface (Figure S6) indicates that these nanostructures correspond to two (111) fluorite planes. Moreover, the contact between the bilayer and the support involves the (111) planes of MgO. Therefore, the interface between the two components can be described as Ce$_{0.5}$Tb$_{0.5}$O$_x$ (111) || MgO (111). Figure 6(b) shows one of these bilayers imaged from top, particularly along the [110] Ce$_{0.5}$Tb$_{0.5}$O$_x$ zone axis, as established from the details of the diffractogram of this area shown in Figure S7(b).

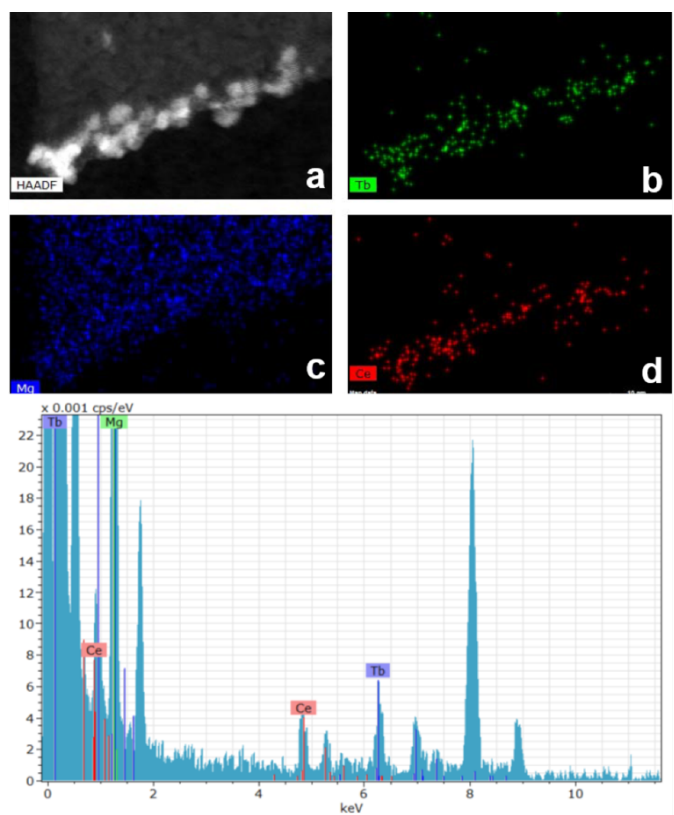

Figure 4. (a) Low mag HAADF-STEM image of Ce$_{0.5}$Tb$_{0.5}$O$_x$ (3% mol.)/MgO; (b-d) EDS Chemical maps of the different elements; (e) EDS spectrum of the whole area in (a).

This diffractogram shows, in addition to the {111} and {200} reflections of the fluorite phase, very weak {111} and {200} MgO spots aligned with the former, which reveals an oriented growth of the bilayer on top of the MgO support.

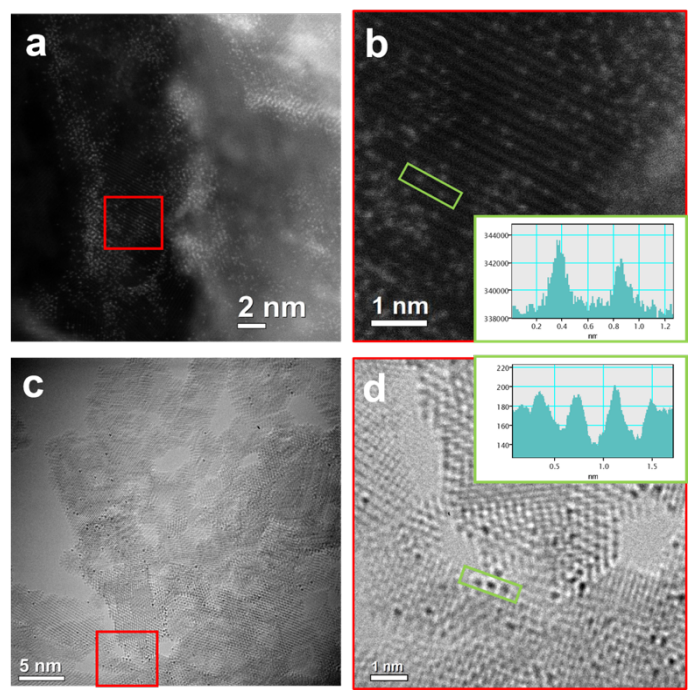

Figure 5. HAADF-STEM (a,b) and HREM (c,d) images of the Ce$_{0.5}$Tb$_{0.5}$O$_x$(3% mol.)/MgO-SRMO sample.

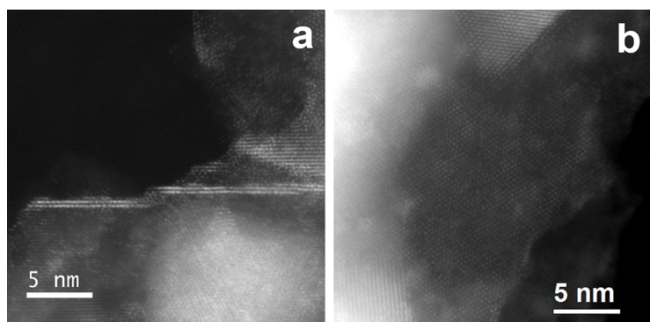

Figure 6. HAADF-STEM image recorded of $Ce_{0.5}Tb_{0.5}O_x$ (3% mol.)/MgO-SRMO sample (a) extended bilayers and (b) these bilayers imaged from top.

To confirm these structural details, a model consisting of two extended (111) fluorite layers deposited in parallel alignment on top of a (111) MgO surface was built and the corresponding HAADF-STEM image simulated, Figure S8. Note how the contrasts in the simulated image, Figure S8(b), match perfectly those observed experimentally. In particular, it is interesting to highlight how the contrasts associated to the atomic columns populated by the lanthanide elements, of large Z, are dominant, whereas those corresponding to the location of the $Mg^{2+}$ cations are hardly observed, in very good agreement with the experimental image.

To clarify the chemical nature of these bilayers, atomically resolved Electron Energy Loss Spectroscopy (EELS) was performed. Figure 7 gathers the results of the analysis of the bilayer in Figure 6. The area of the layer in which the chemical analysis was performed is marked in Figure 7(a). Figures 7(b) through 7(e) show the Annular Dark Field signal, Fig. 7(b), which allows to localize the atomic columns, together with the spectrum images corresponding to the $M_{4,5}$ signals of Ce, Fig. 7(c), and Tb, Fig 7(d), as well as the composite image where the two lanthanide are superimposed, Fig. 7(e). Note first how the two lanthanides are present in the layer, which indicates that they are mixed at atomic level. Moreover, they seem to be distributed in a patch-like fashion with Tb-rich, sub-nanometer sized, regions. In any case it is clear that these nanostructures correspond to a ceria-terbia mixed oxide.

STEM-EELS analysis in the spectrum imaging mode allowed also to investigate the redox state of the lanthanides in the bilayers. To this end, the collection of spectra was first denoised by the Principal Component Analysis (PCA) method. Figure 8 shows the results of this analysis performed on a nanometer sized bilayer. ADF and chemical maps of Ce and Tb of the structure marked on Figure 8(a), are shown in Figures 8(b)-(d). Figure 8(e) corresponds to the composite Ce+Tb signal. Once more, the patch-like distribution of Ce and Tb within the layer is observed. Moreover, taking into account that the fine structure of the $M_{4,5}$ signals of Ce allows differentiating between its +4 and +3 oxidation state,[45] the collection of spectra was analysed by the Non-negative Matrix Factorization Components (NMF) method,[46] to map the distribution of the oxidation states of this element at atomic level, Figures 8(f)-(k).

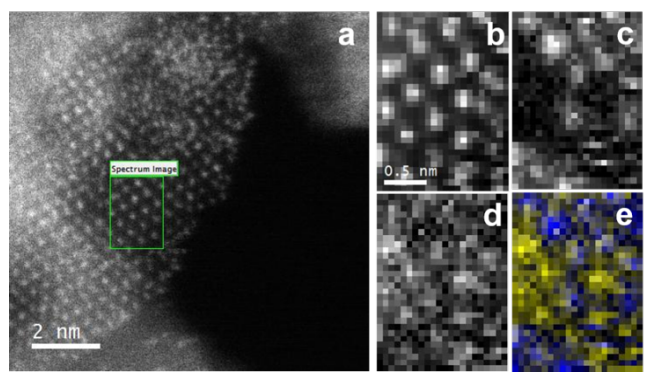

Figure 7. Electron Energy Loss Spectroscopy (EELS) SI experiments of bilayer structures in $Ce_{0.5}Tb_{0.5}O_x$ (3% mol.)/MgO SRMO sample: (a) HR-HAADF-STEM image. The rectangle marks the area where Spectrum Imaging was performed; (b) ADF image of the SI image; (c) Ce map; (d) Tb map; (e) Ce (blue) and Tb (yellow) overlaid maps.

According to this analysis, the spectra of the bilayers contain three major components; two related to Ce ($Ce^{3+}$ and $Ce^{4+}$ species), and only one Tb component which indicates a dominant contribution of $Tb^{3+}$. This result therefore indicates not only mixture between the lanthanides at atomic level but also mixture of electronic states, in the case of Ce. This result would be in agreement with the presence of $Ce^{3+}$ detected by XPS, whose fraction increases up to nearly 25% in this catalyst. STEM-EELS-SI experiments performed on bilayers imaged in the direction parallel to the interface with the MgO support, Figure S9, confirm the presence of both lanthanides in these unique nanostructures.

To determine additional details of the structure of the bilayers, the HAADF-STEM images were analyzed by a procedure involving a combination of smart image denoising, using Undecimated Wavelet Transforms (UWT),[47] and contrast localization by Template Matching methods,[48]

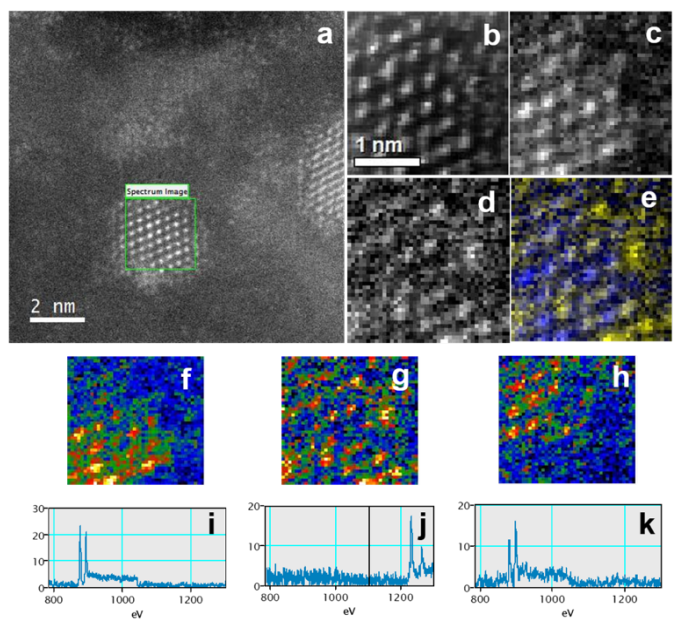

Figure 8. STEM-EELS-SI analysis of the oxidation state of the lanthanides in the bilayers observed in the $Ce_{0.5}Tb_{0.5}O_x$ (3% mol.)/MgO SRMO sample.

Thus, Figures 9(a) and (b) depict the contrasts of a HAADF-STEM image of a layer, recorded in top view along the [110] direction of the fluorite structure, before and after UWT denoising. After background subtraction, Figure 9(c), the position of the contrast maxima was automatically determined using the correlation with a template obtained by averaging the contrasts of different atomic column-like contrasts all over the image, Figure 9(d). After this operation, a map of the X-Y position of the centroids of all the detected atomic columns in the image was automatically generated, Figure 9(e). This analysis reveals certain inhomogeneity in the structure. In fact, at the level of first neighbors, the distribution of atomic columns corresponds to that expected in this case for a [110] zone axis image, but there are evident distortions at larger distances as well as differences in the occupation of atomic columns, in such a way that certain columns appear as missing. Therefore, these extended layers seem to be formed by slightly misaligned domains with fluorite type structure. The segregation of Tb and Ce in small patches, as detected by the analytical tools, maybe responsible of this structural feature. Moreover, this local disorder at medium distances may explain why in the HAADF-STEM images of the bilayers recorded edge-on view, as that in Figure 6(a), the fluorite structure of the mixed ceria-terbia oxide is not clearly revealed.

The structural model obtained after template matching allows also determining the Radial Distribution Function (RDF) of the atomic columns, Figure 9(f). Note that the RDF shows a first maximum at 0.387 nm overlapped with a second around 0.420 nm. These distances are larger than those expected for the Ce-Ce or Tb-Tb distances in the [110] projection of a fluorite structure (0.331 nm and 0.382 nm). This indicates that the growth of the (111) bilayers on top of the (111) MgO planes involves some strain.

Apart from the three type of lanthanide containing nanostructures already described, a fourth type is also present in the SRMO treated catalyst, 3D nanoparticles of medium size, 10-20 nm, Figure 10. As depicted in Figures 10(b) and (c), the composition of these nanoparticles do indeed involve both Ce and Tb. The quantification of the STEM-XEDS spectra recorded on these particles is close to the average one, 51% Ce-49%Tb. It is also quite interesting the extremely sharp interface interface with the MgO, which in this case involves a (100) $Ce_{0.5}Tb_{0.5}O_x$ || (111) MgO contact, Figure 10(e). Note at this respect how in the diffractogram of the image in Figure 10(e), Figure S10(b), the (200) spot of $Ce_{0.5}Tb_{0.5}O_x$ is parallel to the (111) of the support. Another quite interesting feature of the structure of the nanoparticles is the alternation of intensities of the {002} type lattice planes, which is particularly well observed at locations as that marked with a rectangle. This image feature is evidenced both in the intensity profile shown in Figure 10(f) and in the diffractogram of Figure S10(b). Since the contrasts in this imaging mode are dominated by the contribution of the cations, this feature points out to the alternation in space of higher and lower average atomic number planes. Therefore, this observation would suggest the occurrence of alternated Tb-rich and Ce-rich {002} planes.

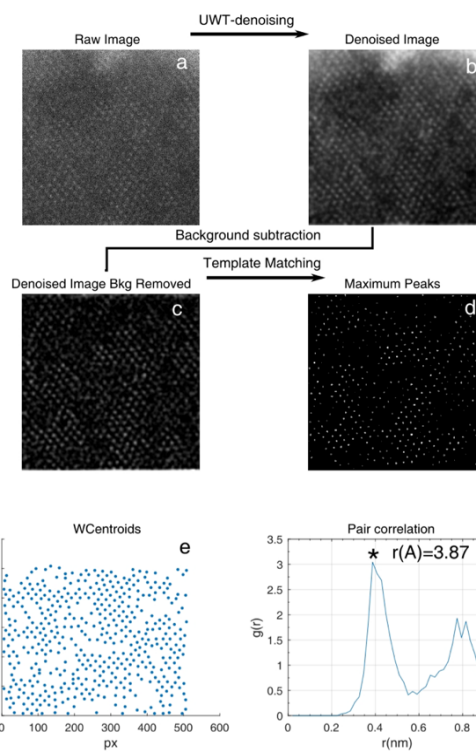

Figure 9. RDF analysis of the HAADF-STEM images of bilayers in $Ce_{0.5}Tb_{0.5}O_x$ (3% mol.)/MgO SRMO.

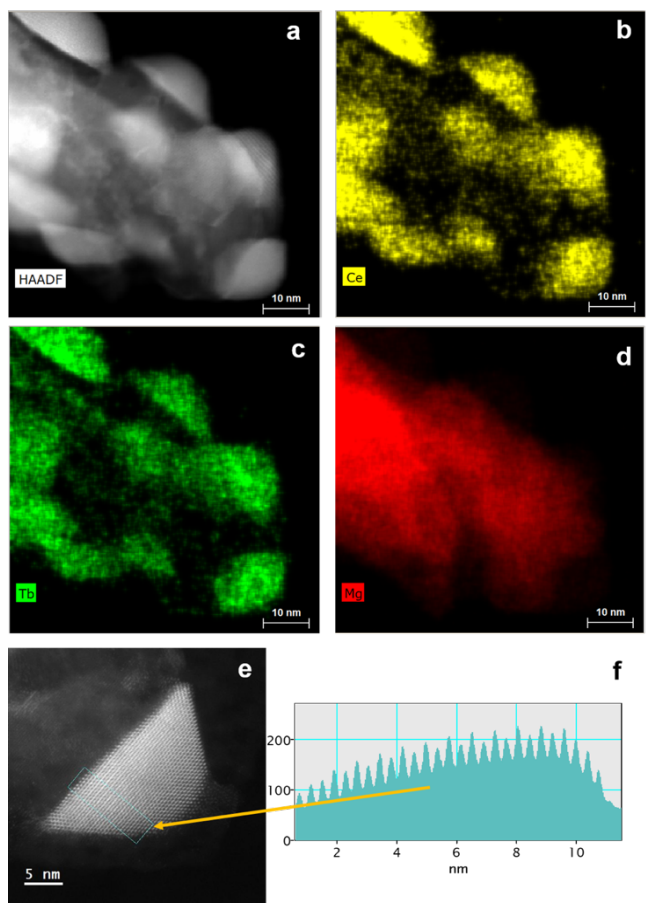

Figure 10. STEM-XEDS image recorded of 3D nanoparticles of medium size, 10-20 nm of $Ce_{0.5}Tb_{0.5}O_x$ (3% mol.)/MgO-SRMO sample.

Nevertheless, A.C. Johnston-Peck et al.[49] have proved that HR STEM-HAADF images of cubic, fully reduced, ceria, i.e. $Ce_2O_3$, show double periodicity {002} fringes in images recorded along the [100] zone axis. This is related to the occurrence of subtle displacements in the position of the Ce atom columns linked to structural relaxation around oxygen vacancies. To check if this was also the case in images recorded along the [110] zone axis, a model of the cubic sesquioxide, C-$Ln_2O_3$, in which the two lanthanides were disordered, was built and the corresponding image calculated, Figure S11. It is clear that also along this zone axis a double {200} periodicity is observed, due to the slight shifts of the cation columns with respect to those in the ideal fluorite.

According to XPS and STEM-EDS data, the composition of these 3D nanoparticles should be close to $Ce_{0.5}Tb_{0.5}O_{1.70}$, since Tb is almost fully reduced and roughly 25% Ce is present as $Ce^{3+}$. This reduction degree, though still far from that of the fully reduced sesquioxide, falls within the oxygen-composition range of the σ-phase (C-$LnO_{1.5+x}$), which spans the O/Ln ratio from 1.5 (bixbyite C-$Ln_2O_3$) up to 1.71 ($\iota$-$Ln_7O_{12}$).[50]

HREM images of these 3D nanoparticles, Figure S12, clearly show the presence of a fluorite superstructure. The superstructure features are observed in the images in the form of low frequency contrast modulations, Figure S12(a), which also reflect in the diffractograms of the HREM images as a set of extra spots located in between those of the fluorite, Figure S12(b). A comparison of the diffractograms in Figures S10(b) and S12(b) with those calculated for the different members of the homologous series of the higher rare earth oxides,[51] along the [110] zone axis, reveals a good match with both the C-$LnO_{1.5+x}$ and the π-$Ln_{16}O_{30}$ phases ($LnO_{1.875}$).

The reduction degree in the π -$Ln_{16}O_{30}$ phase would involve only 24% reduction of the lanthanides and, therefore, nearly half of the Tb in the nanoparticles to be $Tb^{4+}$. From XPS the presence of a small fraction of $Tb^{4+}$ in these nanoparticles cannot be fully disregarded since their size, 15-20 nm, is larger than the depth of analysis of the Tb 3d signal. Nevertheless, since EELS results point out to a dominance of $Tb^{3+}$, the bixbyite type structure seems to be the most likely. In any case, a small $Tb^{4+}$ content cannot be fully disregarded, and in fact it could contribute to the 47% of OSC observed in $Ce_{0.5}Tb_{0.5}O_x$ (3% mol.)/MgO SRMO after reduction at 700ºC; a value higher than that expected from the $Tb^{3+}$ and $Ce^{3+}$ contents detected by XPS, nearly 100% and 23% respectively.

Whatever the exact reduced phase could be, it seems clear that the superstructure features detected both in HAADF-STEM and HREM images can be fully explained without involving disorder-order transitions in the cationic sublattice, as it has been described for ceria-zirconia mixed oxides submitted to SRMO type treatments.[52]

In any case, the spatial distribution and the oxidation states of Ce in these 3D nanoparticles were also investigated by means of STEM-EELS-SI, Figures 11 and S12. Figure 11(a) shows a HAADF-STEM image where one of these studies was performed. Figure 11(b) depicts the ADF image of the area where the SI signal was acquired. The analysis of the whole collection of spectra by PCA and NMF reveals the presence of three independent components, one due to $Ce^{3+}$, another due to $Ce^{4+}$ and a third one related to Tb. The maps of these three components are shown as Figures 11(d) through (f). The intensity of the component in each pixel is color-coded, in such a way that blue represents the lowest intensity and yellow the highest one. Note, first, that both ($Ce^{3+}$ + $Ce^{4+}$) and Tb signals are present throughout the nanoparticle volume, from surface to bulk, which indicates mixture of the two components at the atomic level.

Concerning the oxidation states of Ce, it is clear that both $Ce^{3+}$ and $Ce^{4+}$ are present in the bulk and in the surface. The intensity distribution of $Ce^{3+}$ looks more homogeneous, whereas the signal of $Ce^{4+}$ gets more intense at the thicker parts of the nanoparticle. Moreover, the area covered by the $Ce^{3+}$ signal is slightly smaller than that of $Ce^{4+}$, the intensity of the former being lost at locations close to the interface with the MgO support. These observations suggest the likely presence of a $Ce^{3+}$ shell covering the particle and an accumulation of oxidized Ce at the interfaces with the MgO support. Additional STEM-EELS-SI results included in Figure S13 provide further support to these evidences.

**Methane Oxidation Performance**

Methane conversion *vs* temperature plots included in Figure 12 allow a comparison of the methane oxidation activity between the terbium modified and pure $CeO_2$ supported on magnesia catalysts. Note at this respect that the incorporation of the second lanthanide shifts the conversion plot to the lower temperature range, which points out to an improvement in catalytic performance in this reaction. The improvement is even higher in terms of efficiency in the usage of the lanthanide elements, since the total molar loading in lanthanides of the terbium-modified catalyst is just half that of the catalyst based on pure ceria.

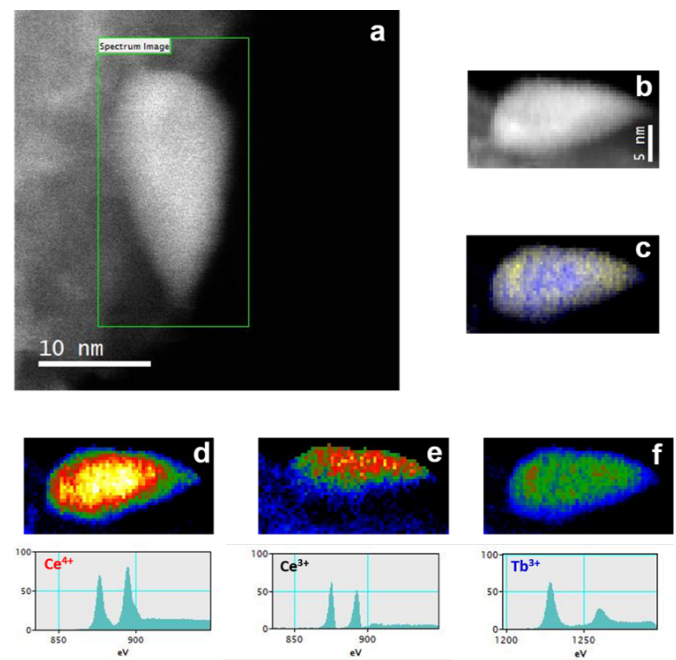

Figure 11. STEM-EELS-SI image recorded on the 3D nanoparticles of medium size 10-20 nm in $Ce_{0.5}Tb_{0.5}O_x$ (3% mol.)/MgO-SRMO.

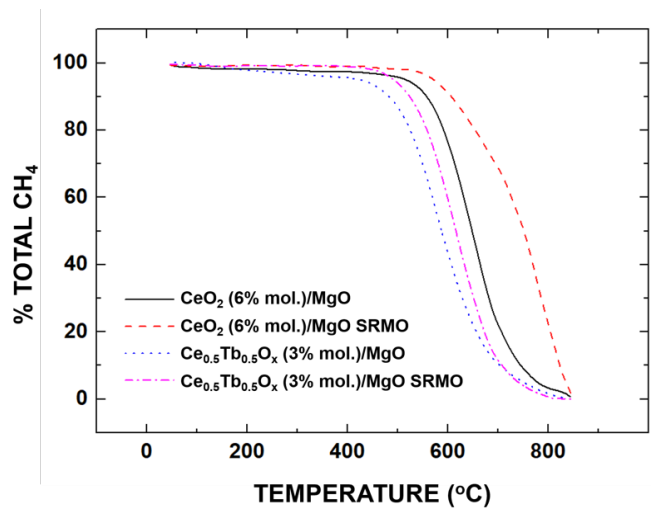

Figure 12. Methane oxidation experiments of the prepared catalysts in comparison with $CeO_2$ (6% mol.)/MgO catalysts.

The enhancement of catalytic performance is also observed in the values at which certain $CH_4$ conversion values are obtained, Table 5. Thus, for 10% and 50% conversions, an 80ºC shift to lower temperatures is observed after addition of Tb. Even at the highest conversions, a 50ºC difference is still observed. This improvement parallels that observed in the redox performance of the catalysts, both in terms of reducibility and OSC.

Though differences in the experimental conditions employed in the catalytic tests preclude a precise comparison with other ceria-based systems, some meaningful comparisons can still be done. Thus, Trovarelli et al.[41] reported a light-off temperature ($T_{50}$) for a bulk-type pure $CeO_2$ oxide of roughly 670ºC, in the order of that observed for $CeO_2$ (6% mol.)/MgO but still 80ºC above that of $Ce_{0.5}Tb_{0.5}O_x$(3% mol.)/MgO. In any case, the two latter behave much better than pure ceria in terms of usage of the lanthanides, especially the catalyst based on mixed ceria-terbia.

Modification of ceria with other lanthanide elements which increase its reducibility, in bulk type catalysts, have been also reported. Thus, R.O. Fuentes et al.[53] have recently found $T_{50}$ values of 570ºC and 530ºC for $Ce_{0.9}Gd_{0.1}O_{2-x}$ and $Ce_{0.9}Pr_{0.1}O_{2-x}$ bulk type catalysts, which are either in the order or below that of the catalyst prepared in this work, but involving amounts of lanthanides nearly two orders of magnitude higher.

A final comparison can be established with a compositionally related ceria-magnesia system, in particular a $Ce_{0.1}Mg_{0.9}O_{1.1}$ catalyst. For this system, M. Chen et al.[54] determined $T_{10}$, $T_{50}$ and $T_{90}$ values of 500ºC, 570ºC and 630ºC.

Therefore, the low or medium temperature behavior of this system is roughly the same or a little worse, at very low temperatures, than that of the $Ce_{0.5}Tb_{0.5}O_x$ (3% mol.)/MgO catalyst here prepared. Nevertheless, even in this case the amount of lanthanide involved is more than three times higher. A second question refers to the effect of high temperature redox aging cycles on catalytic activity. At this respect, it is clear that the SRMO treatment deteriorates, though moderately, the performance of the as-prepared $Ce_{0.5}Tb_{0.5}O_x$ (3% mol.)/MgO catalyst at low and intermediate temperatures. More important, the addition of Tb improves largely the stability, since the increase in the characteristic temperature values ($T_{10}$, $T_{50}$ or $T_{90}$) after SRMO is much larger in the case of the $CeO_2$ (6% mol)/MgO catalyst.

In both systems, $CeO_2$ (6% mol.)/MgO and $Ce_{0.5}Tb_{0.5}O_x$ (3% mol.)/MgO, a large textural change takes place after SRMO, which involves a large drop of both surface area and lanthanide dispersion. As already commented, the small, patch-like, nanoparticles detected in the $Ce_{0.5}Tb_{0.5}O_x$ (3% mol.)/MgO catalyst transform into a complex mixture of nanostructures including bilayers, nano-sized patches and 3D nanoparticles (10-20 nm in size) in $Ce_{0.5}Tb_{0.5}O_x$ (3% mol.)/MgO-SRMO. Though STEM-HAADF revealed a larger fraction of isolated atoms in $Ce_{0.5}Tb_{0.5}O_x$ (3% mol.)/MgO SRMO, this is possibly not enough to compensate the dispersion loss induced by the formation of the other nanostructures, particularly that related to the 3D nanoparticles.

In terms of redox properties, if we recall the $H_2$-TPR and OSC results, the reducibility improves in terms of temperature in the case of $Ce_{0.5}Tb_{0.5}O_x$ (3% mol.)/MgO whereas it remains more or less the same in $CeO_2$ (6% mol.)/MgO. Nevertheless, OSC values decrease slightly in $Ce_{0.5}Tb_{0.5}O_x$ (3% mol.)/MgO after SRMO.

The balance between an improved redox performance and a loss of dispersion due to sintering of support and supported phase results in a moderate loss of activity in both catalytic systems, but compensation works much better in the Tb-modified catalysts, which results in a reasonably good stability against very high temperature treatments. In fact, the temperature of nearly full methane conversion, $T_{90}$, does not change at all in $Ce_{0.5}Tb_{0.5}O_x$ (3% mol.)/MgO whereas it increases by 80ºC in $CeO_2$ (6% mol)/MgO.

Regarding stability, let´s finally point out that none of the papers previously cited provided information about this crucial feature. Therefore, no comparison can be established at this respect.

Summarizing the results of catalytic performance, the $Ce_{0.5}Tb_{0.5}O_x$ (3% mol.)/MgO catalyst prepared in this work performs much better than $CeO_2$ both as bulk and supported on magnesia. In comparison with other bulk materials involving modification of ceria with a second metallic element, such as Zr, Gd or Pr, performances are, in general, similar, particularly at low temperatures. However, in a per-mole of lanthanide terms the $Ce_{0.5}Tb_{0.5}O_x$ (3% mol.)/MgO catalyst behaves comparatively much better, which represents a large improvement in terms of atomic economy in the usage of lanthanides. This is also the case when the comparison is established with bulk-type ceria-magnesia solid solutions involving above 300% higher Ce

Table 5. Comparison of methane oxidation activity values.

| SAMPLE | $T_{10}$ (ºC) | $T_{50}$ (ºC) | $T_{90}$ (ºC) |
|---|---|---|---|
| $CeO_2$ (6% mol.)/MgO * | 560 | 666 | 746 |
| $CeO_2$ (6% mol.)/MgO SRMO * | 606 | 752 | 826 |
| $Ce_{0.5}Tb_{0.5}O_x$(3% mol.)/MgO | 486 | 586 | 696 |
| $Ce_{0.5}Tb_{0.5}O_x$(3% mol.)/MgO SRMO | 528 | 615 | 697 |

* From ref.24

loadings. In terms of stability against high temperature aging treatments, the prepared catalyst suffers only moderate deactivation after a redox cycle involving temperatures above 900ºC but, to the best of our knowledge, there is no data available to provide a comparison with related catalysts.

The whole set of characterization data clearly evidence a high dispersion state of the lanthanide supported phase in the $Ce_{0.5}Tb_{0.5}O_x$ (3% mol.)/MgO catalyst as well as a high degree of interaction between Ce and Tb. Coexistence of both oxidation states, +3 in the case of Tb and a small fraction of Ce, and +4 in the case of most Ce atoms, provides both vacancies which could eventually increase oxygen mobility in the structure and still keep a source of reducible species to substantiate oxygen exchange with reactants. These features reflect both in the redox response and the catalytic performance of the solid.

After SRMO, the interaction state between the two lanthanides is fully retained in the nanostructures observed in the catalyst except, of course, the case of isolated atoms. The appearance of the bilayers and the 3D nanoparticles depicting a bixbyite-like structure may explain the improvement in reducibility observed after SRMO, in terms of a much higher contribution in the $H_2$-TPR of the low temperature range.

The increase in the fraction of isolated atoms occurring after SRMO, though impossible to quantify by STEM techniques, does in any case suggest a larger coverage of the support surface by lanthanide-containing species. This could be also an interesting feature in terms of facilitating further interactions with a metallic supported phase. Note that, although involving a very small fraction of the lanthanide loading, the impact on increasing the probability of interaction with a supported phase is extremely high.

## Conclusions

A magnesia supported mixed ceria-terbia catalyst incorporating a very low loading of the two lanthanides has been prepared by conventional, wet, non-aqueous, incipient impregnation methods. The structure of both the as prepared and high-temperature redox cycled material has been determined in depth by a combination of both macroscopic and atomic scale resolution techniques. These reveal a unique structure, comprising different highly dispersed objects, which span from isolated lanthanide atoms; nano-sized, fluorite-type, lanthanide patches, clusters or particles; lanthanide, fluorite-type, extended bilayers and 3D mixed oxide nanoparticles depicting a bixbyite-like structure with ordered oxygen vacancies. Except, naturally, in the case of isolated atoms, in all these nanostructures the two lanthanides mix together, which ensures their interaction. This interaction translates into a large improvement in the functional behavior, with respect to a similar catalyst based on pure ceria, both in terms of redox properties and catalytic activity in total methane oxidation. Likewise, the prepared $Ce_{0.5}Tb_{0.5}O_x$ (3% mol.)/MgO catalyst depicts a reasonable stability against treatments which involve redox cycling at very high temperatures.

Considering the whole set of structural and functional features, it is clear that the $Ce_{0.5}Tb_{0.5}O_x$ (3% mol.)/MgO catalyst features a number of properties of large interest to be used as support of metallic phases, particularly to prepare methane oxidation catalyst. The high dispersion state of the lanthanides allows a good interaction with the support as well as a proper coverage of its surface, which will facilitate the establishment of further metal-lanthanide phase contacts. In these contacts, the lanthanide containing phase demonstrates an enhanced oxygen exchange capability which is in fact maintained after quite severe temperature treatments. This would guarantee a stable source of oxygen. Moreover, the intrinsic methane oxidation behavior of the prepared catalyst, which competes with that of other ceria-based formulation and clearly improves all of them in terms of efficiency in the usage of the lanthanide, contributes also to considering $Ce_{0.5}Tb_{0.5}O_x$ (3% mol.)/MgO as a promising candidate to prepare active and stable methane oxidation catalysts.

## Conflicts of interest

There are no conflicts to declare.

## Acknowledgements

We acknowledge financial support from FEDER/MINECO Projects (Ref: MAT2017-87579-R and MAT2016-81118-P) and Junta de Andalucía (FQM334, FQM110). The Electron Microscopy data were all acquired at DME SC-ICyT UCA.